\documentclass[aps,twocolumn,prc,showpacs]{revtex4}
\usepackage{mathrsfs}
\usepackage{longtable,lscape}
\usepackage{txfonts}
\usepackage{amssymb}
\usepackage{indentfirst}
\usepackage{graphicx,,booktabs}
\usepackage{color}
\usepackage{amssymb}
\usepackage{epsfig}
\newcommand{\vsig}{\mbox{\boldmath$\sigma$\unboldmath}}
\newcommand{\veps}{\mbox{\boldmath$\epsilon$\unboldmath}}

\begin{document}
\title{$\eta'$ photoproduction on the nucleons in the quark model}
\author{
Xian-Hui Zhong$^{1}$ \footnote {E-mail: zhongxh@ihep.ac.cn} and
Qiang Zhao$^{2,3}$ \footnote {E-mail: zhaoq@ihep.ac.cn}}
\affiliation{ 1)  Department of Physics, Hunan Normal University,
and Key Laboratory of Low-Dimensional Quantum Structures and Quantum
Control of Ministry of Education, Changsha 410081, China }

\affiliation{ 2) Institute of High Energy Physics,
       Chinese Academy of Sciences, Beijing 100049, China
}
\affiliation{3) Theoretical Physics Center for Science Facilities,
Chinese Academy of Sciences, Beijing 100049, China}

\begin{abstract}
A chiral quark-model approach is adopted to study the $\gamma
p\rightarrow \eta'p$ and $\gamma n\rightarrow \eta' n$. Good
descriptions of the recent observations from CLAS and CBELSA/TAPS
are obtained. Both of the processes are governed by $S_{11}(1535)$
and $u$ channel background. Strong evidence of an $n=3$ shell
resonance $D_{15}(2080)$ is found in the reactions, which accounts
for the bump-like structure around $W=2.1$ GeV observed in the total
cross section and excitation functions at very forward angles. The
$S_{11}(1920)$ seems to be needed in the reactions, with which the
total cross section near threshold for the $\gamma p\rightarrow
\eta' p$ is improved slightly. The polarized beam asymmetries show
some sensitivities to $D_{13}(1520)$, although its effects on the
differential cross sections and total cross sections are negligible.
There is no obvious evidence of the $P$-, $D_{13}$-, $F$- and
$G$-wave resonances with a mass around 2.0 GeV in the reactions.
\end{abstract}
\pacs{13.60.Le, 14.20.Gk, 12.39.Jh, 12.39.Fe}
\maketitle

\section{Introduction}

The threshold energy of the $\gamma p\rightarrow \eta' p$ and
$\gamma n\rightarrow \eta' n$ reactions is above the second
resonance region, which might be a good place to extract information
of the less-explored higher nucleon resonances around $2.0$ GeV.
Thus, the study of $\eta'$ photoproduction becomes an interest topic
in both experiment and theory. However, due to the small production
rate  for the $\eta'$ via an electromagnetic probe, it had been a
challenge for experiment to measure the $\eta'$ production cross
section in the photoproduction
reaction~\cite{:1968ke,Struczinski:1975ik,Plotzke:1998ua}.

Theoretical analyses can be found in the literature which were
performed to interpret the old data of $\gamma p\rightarrow \eta'
p$~\cite{:1968ke,Struczinski:1975ik,Plotzke:1998ua}. Zhang \emph{et
al.}~\cite{Zhang:1995uha} first analyzed the old data with an
effective Lagrangian approach, in which the off-shell contributions
from the low-lying resonances in $(1.5\sim 1.7)$ GeV were excluded.
They considered that the main contribution to the photoproduction
amplitude came from $D_{13}(2080)$. Li~\cite{Li:1996wj} and
Zhao~\cite{Zhao:2001kk} also studied the reaction within a
constituent quark model approach. They found the dominance of $S$
wave in the $\eta'$ production, and the off-shell $S_{11}(1535)$
excitation played an important role near the $\eta'$ threshold. They
also predicted that effects of higher resonances in the $n=3$ shell
might be observable in experiment. The dominant role of
$S_{11}(1535)$ was also suggested by Borasoy with the $U(3)$ baryon
chiral perturbation theory~\cite{Borasoy:2001pj}, and Sibirtsev
\emph{et al.} with a hadronic model~\cite{Sibirtsev:2003ng}.
Considering the interferences between $S_{11}(1535)$ and the
background ($t$ channel vector meson exchanges), they gave a
reasonable description of the old data. In 2003 Chiang and Yang
developed a Reggeized model for $\eta$ and $\eta'$ photoproduction
on protons~\cite{Chiang:2002vq}. In this model, the differential
cross section data from~\cite{Plotzke:1998ua} can be well described
by the interference of an $S_{11}$ resonance with a mass in the
range of $(1.932\sim 1.959)$ GeV and the $t$ channel Regge
trajectory exchanges. In 2004 Nakayama and
Haberzett~\cite{Nakayama:2004ek} analyzed the differential cross
section data from~\cite{Plotzke:1998ua} within a relativistic meson
exchange model of hadronic interactions. They predicted that the
observed angular distribution is due to the interference between the
$t$-channel and the nucleon resonances $S_{11}(1650)$ and
$P_{11}(1880)$. Although there are some hints of higher nucleon
resonances in the $\eta'$ photoproduction process, it is not
straightforward to extract them based on the old data with large
uncertainties.

With the rapid development in experiment, recently, high-statistics
and large-angle-coverage data for the $\gamma p\rightarrow \eta' p$
reaction have been reported by the CLAS
Collaboration~\cite{Dugger:2005my,Williams:2009yj} and CBELSA/TAPS
Collaboration~\cite{Crede:2009zzb}, respectively. More recently, the
measurements of the quasi-free photoproduction of $\eta'$ mesons off
nucleons bound in the deuteron were also carried out by the
CBELSA/TAPS Collaboration~\cite{Jaegle:2010jg}. The recent new data
not only provide us a good opportunity to better understand the
reaction mechanism but also allows us to carry out a detailed
investigation of the less-explored higher nucleon resonances.
Motivated by the new high-precision cross-section data obtained by
the CLAS Collaboration~\cite{Dugger:2005my}, Nakayama and
Haberzett~\cite{Nakayama:2005ts} updated their fits and found that
higher resonances with $J=3/2$ might play important roles in
reproducing the details of the measured angular distribution. A bump
structure in the total cross around $W=2.09$ GeV is predicted and
might be caused by $D_{13}(2080)$ and/or $P_{13}(2100)$. In the
quark model Li~\cite{Li:1996wj} and Zhao~\cite{Zhao:2001kk} also
found a bump structure around $W= 2.1$ GeV ($E_{\gamma}\simeq2.0$
GeV) in the cross section by analyzing the old data. This structure
comes from the $n=3$ terms in the harmonic oscillator basis. The
later higher-precision free proton data from the CLAS
Collaboration~\cite{Dugger:2005my,Williams:2009yj} indeed show a
broad bump structure in the cross section around $W=2.1$ GeV. This
structure seems to also appear in the very recent quasi-free proton
data and the data for inclusive quasi-free $\gamma d\rightarrow (np)
\eta'$ process~\cite{Jaegle:2010jg}.

To clarify the structures from the above analyses and observations,
we present a systemic analysis of the recent experimental data for
$\gamma p\rightarrow p\eta'$ and $\gamma n\rightarrow \eta 'n$ in
the framework of a chiral quark model as an improvement of the
previous studies~\cite{Li:1996wj,Zhao:2001kk}. The chiral quark
model has been well developed and widely applied to meson
photoproduction reactions
~\cite{qk2,qkk,Li:1997gda,qkk2,qk3,qk4,qk5,Li:1995vi,Li:1998ni,
Saghai:2001yd,He:2008ty,He:2009zzi}. The details about the model can
be found in \cite{Li:1997gda,qk3}. Recently, we applied this model
to study $\eta$ photoproduction on the free and quasifree
nucleons~\cite{Zhong:2011ti}. Good descriptions of the observations
were obtained. In this work, we  extend this approach to $\eta'$
photoproduction. Given that the $\eta'$ and $\eta$ are mixing states
of flavor singlet and octet in the SU(3) flavor symmetry, we expect
that some flavor symmetry relation can be applied to these two
channels as a constraint on the model parameters. Moreover, since
$\eta'$ production has a higher threshold, the determinations of the
low-lying resonances in $(1.5\sim 1.7)$ GeV in the $\eta$
photoproduction would be useful for estimating their off-shell
contributions in the $\eta'$ photoproduction.

Similar to the $\eta$ production, an interesting difference between
$\gamma p\rightarrow \eta' p$ and  $\gamma n\rightarrow \eta' n$ is
that in the $\gamma p$ reactions, contributions from states of
representation $[70,^48]$ will be forbidden by the Moorhouse
selection rule~\cite{Moorhouse:1966jn} in the SU(6)$\otimes$O(3)
symmetry. As a consequence, only states of $[56,^28]$ and $[70,^28]$
would contribute to $\gamma p\rightarrow \eta' p$. In contrast, all
the octet states can contribute to the $\gamma n$ reactions. In
another word, more states will be present in the $\gamma n$
reactions. Therefore, a combined study of the $\eta'$ meson
photoproduction on the proton and neutron should provide some
opportunities for disentangling the role played by intermediate
baryon resonances.

The paper is organized as follows. In Sec.~\ref{fram}, a brief
introduction of the chiral quark model approach is given. The
numerical results are presented and discussed in Sec.~\ref{cy}.
Finally, a summary is given in Sec.~\ref{summ}.

\section{framework}\label{fram}

In the chiral quark model, the $s$- and $u$-channel transition
amplitudes for pseudoscalar-meson photoproduction on the nucleons
have been worked out in the harmonic oscillator basis in
Ref.~\cite{Li:1997gda}. The $t$-channel contributions from vector
meson exchange are not considered in this work. If a complete set of
resonances are included in the $s$ and $u$ channels, the
introduction of $t$-channel contributions might result in double
counting~\cite{Dolen:1967jr,Williams:1991tw}.

It should be remarked that the amplitudes in terms of the harmonic
oscillator principle quantum number $n$ are the sum of a set of
SU(6) multiplets with the same $n$. To see the contributions of
individual resonances, we need to separate out the
single-resonance-excitation amplitudes within each principle number
$n$ in the $s$-channel. Taking into account the width effects of the
resonances, the resonance transition amplitudes of the $s$-channel
can be generally expressed as \cite{Li:1997gda}
\begin{eqnarray}
\mathcal{M}^s_R=\frac{2M_R}{s-M^2_R+iM_R
\Gamma_R}\mathcal{O}_Re^{-(\textbf{k}^2+\textbf{q}^2)/6\alpha^2},
\label{stt}
\end{eqnarray}
where $\sqrt{s}=E_i+\omega_\gamma$ is the total energy of the
system, $\alpha$ is the harmonic oscillator strength, $M_R$ is the
mass of the $s$-channel resonance with a width
$\Gamma_R(\mathbf{q})$, and $\mathcal{O}_R$ is the separated
operators for individual resonances in the $s$-channel. In the
Chew-Goldberger-Low-Nambu (CGLN)
parameterization~\cite{Chew:1957tf}, the transition amplitude can be
written with a standard form:
\begin{eqnarray}
\mathcal{O}_R&=&i f^R_1 \vsig \cdot \veps+f^R_2 \frac{(\vsig \cdot
\mathbf{q})\vsig\cdot (\mathbf{k}\times
\veps)}{|\mathbf{q}||\mathbf{k}|}\nonumber\\
&& +if^R_3\frac{(\vsig \cdot \mathbf{k}) (\mathbf{q}\cdot
\veps)}{|\mathbf{q}||\mathbf{k}|}+if^R_4\frac{(\vsig \cdot
\mathbf{q}) (\mathbf{q}\cdot \veps)}{|\mathbf{q}|^2},
\end{eqnarray}
where $\vsig$ is the spin operator of the nucleon, $\veps$ is the
polarization vector of the photon, and $\mathbf{k}$ and $\mathbf{q}$
are incoming photon and outgoing meson momenta, respectively.

The $\mathcal{O}_R$ for the $n\leq 2$ shell resonances have been
extracted in \cite{Li:1997gda}.  For the $n=3$ shell resonances are
just around the $\eta'$ production threshold, which might play
important roles in the reaction. Thus, in this work we can not treat
them as degenerate any more. Their transition amplitudes,
$\mathcal{O}_R$, for $S_{11}$, $D_{13}$, $D_{15}$, $G_{17}$ and
$G_{19}$ waves are derived in the SU(6)$\otimes$O(3) symmetric quark
model limit, which have been given in Tab.~\ref{CGLN}. The
$g$-factors that appear in Tab.~\ref{CGLN} can be extracted from the
quark model in the SU(6)$\otimes$O(3) symmetry limit, and are
defined by
\begin{eqnarray}
g_3^v &\equiv& \langle N_f|\sum_{j}e_jI_j\sigma_{jz}|N_i\rangle,\\
g_3^s &\equiv& \langle N_f|\sum_{j} e_jI_j|N_i\rangle,\\
g_2^s &\equiv& \langle N_f|\sum_{i\neq j}e_jI_i\vsig_i\cdot
\vsig_j|N_i\rangle/3,\\
g_2^v&\equiv&\langle N_f|\sum_{i\neq j} e_jI_i(\vsig_i\times
\vsig_j)_z|N_i\rangle/2,\\
g_2^{v'}&\equiv&\langle N_f|\sum_{i\neq
j}e_jI_i\sigma_{iz}|N_i\rangle,
\end{eqnarray}
where $|N_i\rangle$ and $|N_f\rangle$ stand for the initial and
final states, respectively, and $I_j$ is the isospin operator, which
has been defined in \cite{Li:1997gda}. For the $\eta$ and $\eta'$
production, the isospin operator is $I_j=1$.

From Tab.~\ref{CGLN} we can see that the $n=3$ resonance amplitudes
$f^R_i(i=1,2,3,4)$ for $S$ and $D$ waves contain two terms, which
are in proportion to $x^2$ and $x^3$, respectively. The term
$\mathcal{O}(x^3)$ is a higher order term in the amplitudes for
$x\equiv |\mathbf{k}||\mathbf{q}|/(3\alpha^2)\ll 1$. For the
$G_{17}$ and $G_{19}$ waves, their amplitudes only contain the high
order term $\mathcal{O}(x^3)$, thus their contributions to the
reactions should be small in the $n=3$ shell resonances. Comparing
the resonance amplitudes $f^R_i(i=1,2,3,4)$ for $D_{13}$ with those
for $D_{15}$, we find that
\begin{eqnarray}
\left|f^R_1[D_{15}(n=3)]\right|&>&\left|f^R_1[D_{13}(n=3)]\right|P_3'(\cos\theta),\label{ampd}\\
\left|f^R_i[D_{15}(n=3)]\right|&>&\left|f^R_i[D_{13}(n=3)]\right|\ \
\ \ (i=2,3,4),
\end{eqnarray}
for the $\eta'$ and $\eta$ photoproduction processes. The amplitude
$f^R_1$ for $D_{13}$ is reaction angle independent, while the
$f^R_1$ for $D_{15}$ depends on the reaction angle $\theta$ (i.e.
$\propto P_3'(\cos\theta)$). According to Eq.~\ref{ampd}, at very
forward and backward angles [i.e. $\cos \theta \simeq \pm 1$] we
obtain
\begin{eqnarray}
\left|f^R_1[D_{15}(n=3)]\right|_{\cos\theta\simeq \pm 1}&>& 6
\left|f^R_1[D_{13}(n=3)]\right|.
\end{eqnarray}
It shows that the magnitude of $f^R_1$ at very forward and backward
angles for $D_{15}$ is about an order larger than that of $D_{13}$.
Thus, the $D_{15}$ partial wave is the main contributor to the
$\eta'$ and $\eta$ photoproduction processes in the $n=3$ shell
resonances. At very forward and backward angle regions, the angle
distributions might be sensitive to the $D_{15}$ partial wave. We
note that due to lack of experimental information and high density
of states above 2 GeV, different representations that contribute to
the same partial wave quantum number in the $n=3$ shell are treated
degenerately as one state as listed in Tab.~\ref{CGLN}.

\begin{widetext}
\begin{center}
\begin{table}[ht]
\caption{ CGLN amplitudes for $s$-channel resonances of the $n=3$
shell in the SU(6)$\otimes$O(3) symmetry limit. We have defined
$A\equiv(\frac{\omega_m}{E_f+M_N}+1)|\mathbf{q}|$,
$x\equiv\frac{|\mathbf{k}| |\mathbf{q}|}{3\alpha^2}$, $P_l'(z)\equiv
\frac{\partial P_l(z)}{\partial z}$, $P_l''(z)\equiv\frac{\partial^2
P_l(z)}{\partial z^2}$, $g_1\equiv g_3^v-\frac{1}{8}g_2^v$,
$g_2\equiv g_3^v-\frac{1}{8}g_2^{v'}$ and $g_3\equiv
g_3^s-\frac{1}{8}g_2^s$. $\omega_\gamma$, $\omega_m$ and $E_f$ stand
for the energies of the incoming photon, outgoing meson and final
nucleon, respectively, $m_q$ is the constitute $u$ or $d$ quark
mass, $1/\mu_q$ is a factor defined by $1/\mu_q=2/m_q$, and $P_l(z)$
is the Legendre function with $z=\cos\theta$.} \label{CGLN}
\begin{tabular}{|c|c|c|c|c|c|c|c|c|c| }\hline\hline
    & $f^R_1$ \ \ & $f^R_2$ & $f^R_3$ & $f^R_4$  \\
\hline $S_{11}$&$-\frac{i}{36}
\frac{\omega_m \omega_\gamma}{\mu_q}(g_2+\frac{k}{2m_q}g_1)x^2$&  & &\\
    & +$\frac{i}{60}(g_1 \frac{k}{m_q}+2g_2)Ax^3$ & 0&0 & 0      \\
\hline $D_{13}$& $\frac{i}{90}\frac{\omega_m
\omega_\gamma}{\mu_q}(g_2+\frac{k}{2m_q}g_1)x^2$&
$\frac{i}{180}\frac{\omega_m
\omega_\gamma^2}{\mu_qm_q}g_1x^2P_2'(z)-\frac{i}{105}$ &
&  $-\frac{i}{90}\frac{\omega_m \omega_\gamma}{\mu_qm_q}g_2x^2P_2''(z)+\frac{i}{420}Ax^3$     \\
        & $-\frac{i}{60}(g_1 \frac{k}{m_q}+2g_2)Ax^3$& $\frac{k}{m_q}(g_1+g_3/2)Ax^3P_2'(z)$ &0
        & $[14g_2-(g_1-g_3)\frac{k}{m_q}]P_2''(z) $     \\\hline
$D_{15}$&$\{-\frac{i}{90}\frac{\omega_m
\omega_\gamma}{\mu_q}(g_2+\frac{k}{2m_q}g_1)x^2+\frac{i}{105}$
&$-\frac{i}{180}\frac{\omega_m
\omega_\gamma^2}{\mu_qm_q}g_1x^2P_2'(z)+\frac{i}{420}$
&$-\frac{i}{90}\frac{\omega_m
\omega_\gamma}{\mu_q}g_2x^2P_3''(z)+\frac{i}{420}$
& $\frac{i}{90}\frac{\omega_m \omega_\gamma}{\mu_q}g_2x^2P_2''(z)-\frac{i}{420}$       \\
        &$[(g_1-\frac{1}{2}g_3)\frac{k}{m_q}+g_2]Ax^3\}P_3'(z)   $
        & $\frac{k}{m_q}(5g_1-3g_3)Ax^3P_2'(z)$
        &$[4g_2-(g_1-g_3)\frac{k}{m_q}]Ax^3P_3''(z)$ & $[4g_2-(g_1-g_3)\frac{k}{m_q}]Ax^3P_2''(z)$
        \\\hline
$G_{17}$&
$\frac{-i}{1890}[(4g_1+5g_3)\frac{k}{m_q}+18g_2]Ax^3P_3'(z)$
&$\frac{-i}{210}(8g_2-g_1\frac{k}{m_q})Ax^3P_4'(z)$&
$\frac{i}{1890}[(g_1-g_3)\frac{k}{m_q}-18g_2]Ax^3P_3''(z)$ &
$\frac{-i}{1890}[(g_1-g_3)\frac{k}{m_q}-18g_2]Ax^3P_4''(z)$
\\\hline
$G_{19}$&$i\frac{2k}{945m_q}(g_1-g_3)Ax^3P_5'(z)$    & $i\frac{k}{378m_q}(g_1-g_3)Ax^3P_4'(z)$
&$-i\frac{k}{1890m_q}(g_1-g_3)Ax^3P_5''(z)$& $i\frac{k}{1890m_q}(g_1-g_3)Ax^3P_4''(z)$    \\
\hline
\end{tabular}
\end{table}
\end{center}
\end{widetext}

\begin{table}[ht]
\caption{The $g$-factor in the amplitudes.}\label{gfac}
\begin{tabular}{|c|ccccccccccccccc| }\hline\hline
reaction &$g_3^v$& &$g_3^s$& &$g_2^s$ &&$g_2^v$ &&$g_2^{v'}$&&
$g_1$&& $g_2$&&$g_3$
\\\hline
$\gamma p\rightarrow \eta' (\eta)
p$&$1$&&$1$&&$0$&&$0$&&$0$& &1&&1&&1 \\
$\gamma n\rightarrow \eta' (\eta)
n$&$-\frac{2}{3}$&&$0$&&$-\frac{2}{3}$&&$0$&&
$-\frac{2}{3}$&&$-\frac{2}{3}$&&$-\frac{3}{4}$&&$\frac{1}{12}$\\\hline
\end{tabular}
\end{table}

Finally, the physical observables, differential cross section and
photon beam asymmetry, are given by the following standard
expressions~\cite{Walker:1968xu}:
\begin{eqnarray}
\frac{d\sigma}{d\Omega}&=&\frac{\alpha_e\alpha_{\eta'}(E_i+M_N)(E_f+M_N)}{16s
M_N^2}\frac{1}{2}\frac{|\mathbf{q}|}{|\mathbf{k}|}\sum^4_{i=1}|H_i|^2,\\
\Sigma &=&2\mathrm{Re}(H_4^*H_1-H_3^*H_2)/\sum^4_{i=1}|H_i|^2,
\end{eqnarray}
where the helicity amplitudes $H_i$ can be expressed by the CGLN
amplitudes $f_i$~\cite{Walker:1968xu,Fasano:1992es}.

\section{CALCULATIONS AND ANALYSIS} \label{cy}

\subsection{Parameters}

In our previous work, we have studied $\eta$ photoproduction off the
quasi-free neutron and proton from a deuteron target, where the
masses, widths and coupling strength parameters $C_R$ of the $n\leq
2$ shell resonances had been determined~\cite{Zhong:2011ti}. In this
work, the same parameter set is adopted. For the $n=3$ shell
resonances, $S_{11}$, $D_{13}$, $D_{15}$, $G_{17}$ and $G_{19}$
waves, their transition amplitudes, $\mathcal{O}_R$, have been
derived in the SU(6)$\otimes$O(3) symmetric quark model limit, which
are given in Tab.~\ref{CGLN}. The various $g$-factors in these
amplitudes for $\eta'$ photoproduction on the nucleons have been
derived in the SU(6)$\otimes$O(3) symmetry limit, which are listed
in Tab.~\ref{gfac}. Their resonance parameters are determined by the
experimental data. The determined mass and width for $D_{15}$ are
$M\simeq2080$ MeV and $\Gamma\simeq80$ MeV, respectively, while the
determined mass and width of $S_{11}$ are $M\simeq1920$ MeV and
$\Gamma\simeq90$ MeV. It should be pointed out that the reactions
are insensitive to the masses and widths of $G$- and $D_{13}$- wave
states in the $n=3$ shell. Thus, in the calculation we roughly take
their mass and width with $M=2100$ MeV and $\Gamma=150$ GeV,
respectively.

There are two overall parameters, the constituent quark mass $m_q$
and the harmonic oscillator strength $\alpha$, from the transition
amplitudes. In the calculations we adopt the standard values in the
the quark model, $m_q=330$ MeV and $\alpha^2=0.16$ GeV$^2$.

To take into account the relativistic effects, the commonly applied
Lorentz boost factor is introduced in the resonance amplitude for
the spatial integrals~\cite{qkk}, which is
\begin{eqnarray}
\mathcal{O}_R(\textbf{k},\textbf{q})\rightarrow
\gamma_k\gamma_q\mathcal{O}_R(\gamma_k\textbf{k}, \gamma_q\textbf{q}
),
\end{eqnarray}
where $\gamma_k=M_N/E_i$ and $\gamma_q=M_N/E_f$.

The $\eta' NN$ coupling is a free parameter in the present
calculations and to be determined by the experimental data. In the
present work the overall parameter $\eta' NN$ coupling
$\alpha_{\eta'}$ is set to be the same for both $\gamma n\rightarrow
\eta' n$ and $\gamma p\rightarrow \eta' p$. The fitted value
$g_{\eta' NN}\simeq 1.86$ (i.e. $\alpha_{\eta'}\equiv g^2_{\eta'
NN}/4\pi=0.275$) is in agreement with that in
Ref.~\cite{Nakayama:2005ts}, where the upper limit of $g_{\eta' NN}$
was suggested to be $g_{\eta' NN}\lesssim 2$. In our previous work
we determined the $\eta NN$ coupling, i.e. $g_{\eta NN}\simeq
2.13$~\cite{Zhong:2011ti}. This allows us to examine the
$\eta-\eta'$ mixing relation for their non-strange components
production,
\begin{eqnarray}
\tan \phi_P=\frac{g_{\eta' NN}}{g_{\eta NN}} \ ,
\end{eqnarray}
which gives $\phi_P\simeq 41.2^\circ$. This value is within the
range of $\phi_P=\theta_P+\arctan\sqrt{2}\simeq 34^\circ\sim
44^\circ$, where $\theta_P\simeq -20^\circ\sim -10^\circ$ is the
flavor singlet and octet mixing angle. The favored value for
$\phi_P$ implies a flavor symmetry between the $\eta$ and $\eta'$
production.

Since the single resonance excitation amplitudes can be separated
out for $n\le 2$ shells, the $\eta'N^*N$ coupling form factor in
principle can be extracted by taking off the EM helicity amplitudes.
The expressions are similar to those extracted in $\eta$ meson
photoproduction~\cite{Zhong:2011ti} apart from the overall $g_{\eta'
NN}$ coupling constant. For higher excited states in $n=3$, due to
the lack of information about their EM excitation amplitudes and
high density of states above the 2 GeV mass region, we treat all
SU(6) multiplets that contribute to the same quantum number in $n=3$
to be degenerate. In this sense, the partial waves in
Tab.~\ref{CGLN} are collective amplitudes from both {\bf 56} and
{\bf 70} representations.

\begin{widetext}
\begin{center}
\begin{figure}[ht]
\centering \epsfxsize=15.0 cm \epsfbox{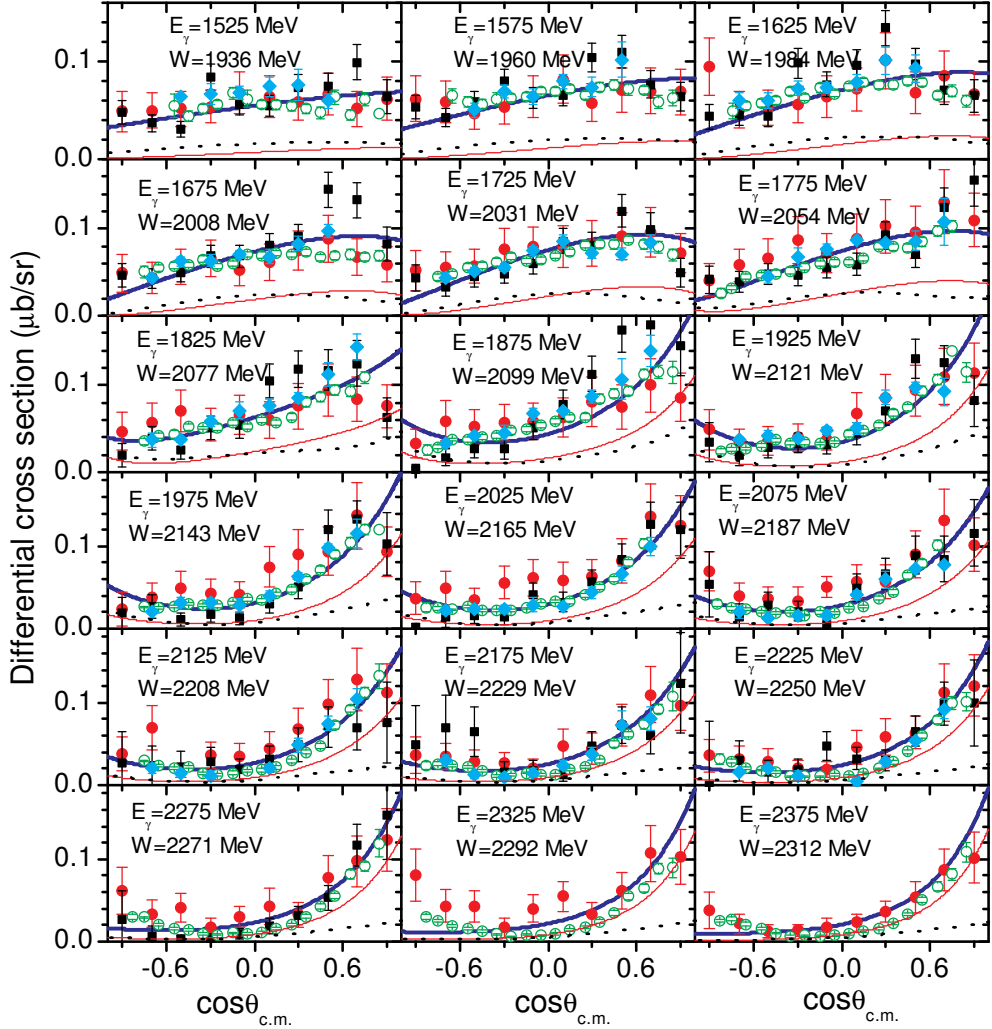} \caption{(Color
online)Differential cross sections for the $\eta'$ photoproduction
off the free proton at various beam energies. The data are taken
from~\cite{Crede:2009zzb} (solid circles), ~\cite{Williams:2009yj}
(open circles), ~\cite{Dugger:2005my} (diamonds). The quasi-free
data from~\cite{Jaegle:2010jg} (squares) are also included. The bold
solid curves stand for the full model calculations. The thin solid
and dotted curves stand for the results without $S_{11}(1535)$ and
background $u$ channel contributions, respectively. }\label{fig-df}
\end{figure}
\end{center}
\end{widetext}

\subsection{$\gamma p\rightarrow \eta' p$ }

\begin{figure}[ht]
\centering \epsfxsize=8.5 cm \epsfbox{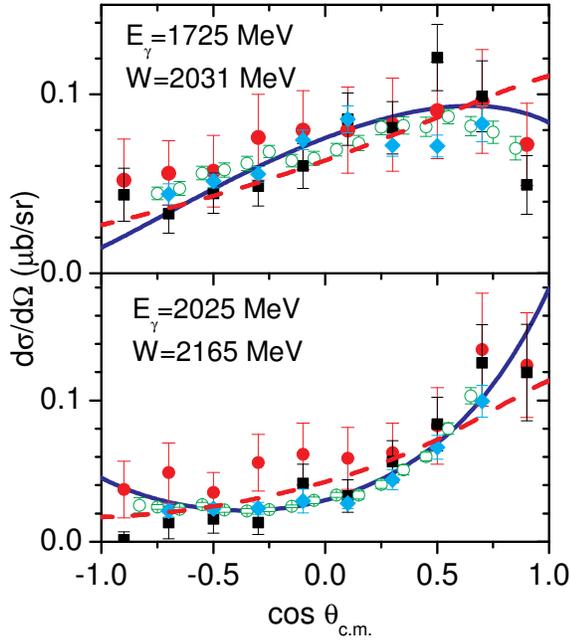} \caption{(Color
online) Same as Fig.~\ref{fig-df}. The dashed curves stand for the
results without $D_{15}(2080)$.}\label{fig-df2}
\end{figure}

\begin{figure}[ht]
\centering \epsfxsize=8.5 cm \epsfbox{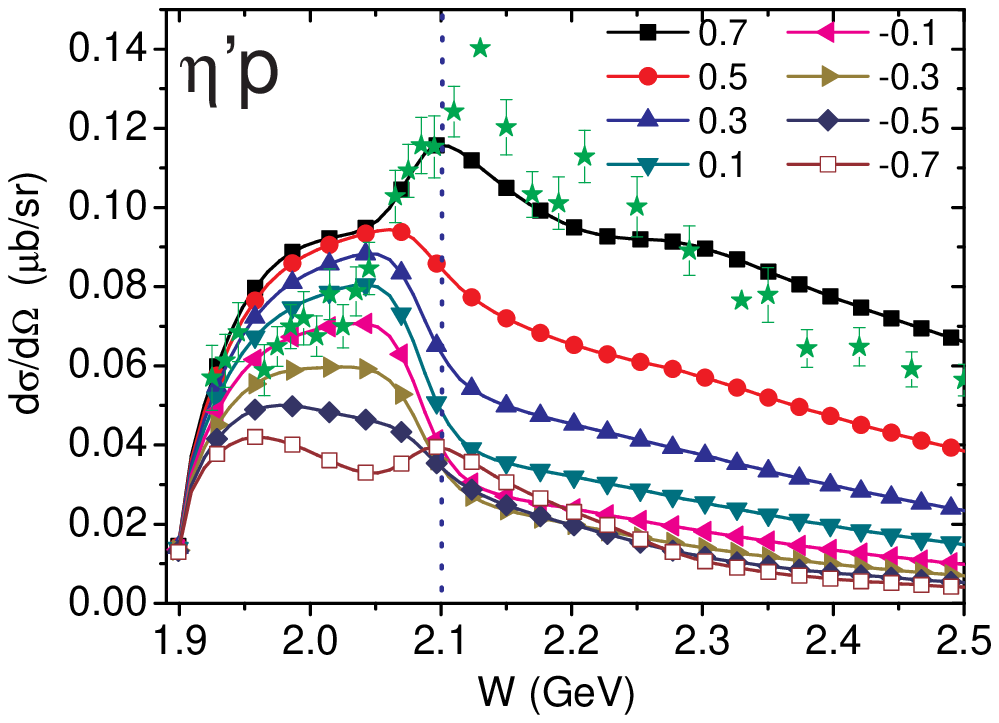}\caption{(Color
online) Fixed-angle excitation functions for $\gamma p\rightarrow
\eta' p$ as a function of center mass energy $W$ for eight
$\cos\theta$, which have been labeled on the plot. The stars stand
for the data from~\cite{Williams:2009yj} for $\cos\theta=0.7$.
}\label{fig-excit}
\end{figure}

The chiral quark model studies of $\gamma p\to \eta' p$ have been
carried out in Refs.~\cite{Li:1996wj,Zhao:2001kk}, where a bump
structure around $E_\gamma=2$ GeV is found arising from the $n=3$
terms in the harmonic oscillator basis. However, which partial wave
contributes to this structure can not be studied in detail since
only a few datum points were available at that time. The improvement
of the experimental situations not only gives us a good opportunity
to better understand the $\gamma p\to \eta' p$ process, but also
allows us to carry out a detailed investigation of the resonances in
the higher mass region.

In Fig.~\ref{fig-df}, we have plotted the differential cross
sections. It shows that our calculations are in good agreement with
the data from threshold up to $E_{\gamma}\simeq 2.4$ GeV.
$S_{11}(1535)$ plays a dominant role in the reaction, switching off
its contributions the differential cross sections are underestimated
drastically. The important role of $S_{11}(1535)$ in the $\gamma
p\to \eta' p$ is also predicted in the previous chiral quark model
study~\cite{Li:1996wj,Zhao:2001kk} and the hadronic model study with
the exchange of vector
mesons~\cite{Sibirtsev:2003ng,Nakayama:2005ts}. It should be
mentioned that the $S_{11}(1535)$ is treated as a mixed state by the
mixing of $[70, ^28]$ and $[70, ^48]$~\cite{Zhong:2011ti}, where the
mixing angle is in agreement with the recent study~\cite{An:2011sb}.

Furthermore, the $u$ channel plays an important role in the
reactions as well. The dotted curves in Fig.~\ref{fig-df} show that
without the contributions of the $u$ channel, the cross sections
will be underestimated significantly. It should be pointed out that
the forward peaks in the differential cross sections are mainly
caused by the $u$ channel backgrounds. The crucial role of
non-resonant background contributions in the $\gamma p\to \eta' p$
is also predicted in Refs.~\cite{Sibirtsev:2003ng,Nakayama:2005ts},
where the $t$ channel vector meson exchanges are the main
non-resonant contributions. In this work, the $t$ channel
contributions are not considered. Since a complete set of resonances
in the $s$ and $u$ channels is included and the $\eta'$ threshold is
rather high, we do not include the $t$ channel exchanges to avoid
the double counting
problem~\cite{Dolen:1967jr,Williams:1991tw,Li:1995vi}.

It is interesting to see that $D_{15}(2080)$ in the $n=3$ shell
plays a crucial role in the reaction. It causes a shape change in
the differential cross section around the $D_{15}(2080)$ mass region
(i.e. $E_\gamma\simeq 1.8$ GeV). In Fig.~\ref{fig-df2} we
demonstrate the interfering effects of $D_{15}(2080)$ by switching
off it in the differential cross section below and above the mass of
$D_{15}(2080)$. It could be obvious evidence of $D_{15}(2080)$ in
the $\gamma p\to \eta' p$ process. We have noted that another
$D$-wave state, $D_{13}(2080)$, was predicted to have significant
effects on the reaction in ~\cite{Zhang:1995uha, Nakayama:2005ts}.
However, in our approach the contributions of the $D$-wave states
with $J^{P}=3/2^-$ in the $n=3$ shell are negligible. The dominant
features of $D_{15}$ in the $D$ wave states can be well understood
from their amplitudes, which has been discussed in Sec.~\ref{fram}.
The amplitude $f^R_1$ for $D_{15}$ is in proportion to
$P_3'(\cos\theta)=(15\cos^2\theta-3)/2$, which can naturally explain
the strong effects of $D_{15}(2080)$ on the deferential cross
sections at forward and backward angles (i.e. $\cos\theta\simeq\pm
1$).

The effects of $D_{15}(2080)$ can be expected  in $\gamma
p\rightarrow \eta p$ taking into account the mixing between $\eta'$
and $\eta$. A recent quark model study of $\eta$ photoproduction in
the high energy region has reported  effects from
$D_{15}(2080)$~\cite{He:2008ty,He:2009zzi}. Evidence of
$D_{15}(2080)$ was also found by a partial wave analysis of the
$\eta$ photoproduction data from CB-ELSA~\cite{Crede:2003ax}  in the
Bonn-Gatchina (BnGa) model~\cite{Anisovich:2005tf}. Its contribution
to  $\gamma p\rightarrow K^+\Lambda$ was also
reported~\cite{Anisovich:2011ye}. Our analysis of the partial wave
amplitudes in Sec.~\ref{fram} also suggests that the $D_{15}$
amplitude plays a dominant role in the $n=3$ shell $D$ wave states
in $K$ photoproduction.

We also mention that $P_{13}(1900)$ can slightly enhances the
differential cross sections around the $\eta'$ production threshold
as found in the previous studies as
well~\cite{Zhao:2001kk,Chiang:2002vq}. It has a similar behavior to
the $u$ channel, although its contribution is much less than that of
the $u$ channel. It could be difficult to identify $P_{13}(1900)$ in
the $\gamma p\to \eta' p$ process in the cross section measurement.
Similar conclusion is found in Ref.~\cite{Chiang:2002vq}. In our
study, contributions from other individual resonances are rather
small, and we do not find obvious signals for states, such as higher
$S_{11}$ states.

\begin{figure}[ht]
\centering \epsfxsize=8.5 cm \epsfbox{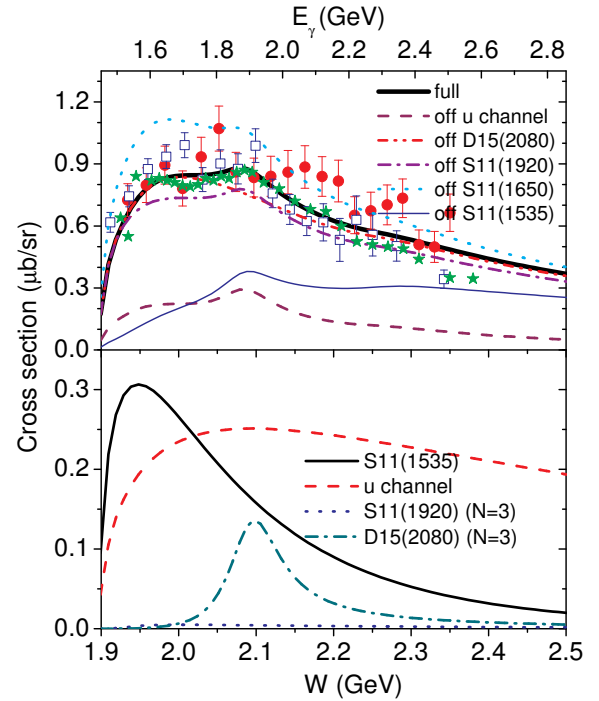} \caption{(Color
online) The cross sections for the $\eta'$ photoproduction off the
free proton. The data are taken from~\cite{Crede:2009zzb} (solid
circles), ~\cite{Williams:2009yj} (stars). The quasi-free data
from~\cite{Jaegle:2010jg} (squares) are also included. In the upper
panel the bold solid curve corresponds to the full model result,
while the thin solid, dotted, dash-dotted, dash-dot-dotted and
dashed curves are for the results by switching off the contributions
from $S_{11}(1535)$, $S_{11}(1650)$, $S_{11}(1920)$, $D_{15}(2080)$
and $u$ channel, respectively. In the lower panel the partial cross
sections for the main contributors are indicated explicitly by
different legends. }\label{fig-crp}
\end{figure}

In Fig.~\ref{fig-excit} we have plotted the fixed-angle excitation
functions for $\gamma p\rightarrow \eta' p$. Our calculations show
that at very forward (e.g. $\cos\theta=0.7$) and backward scattering
angles (e.g. $\cos\theta=-0.7$), there is a bump around $W=2.1$ GeV.
At forward angles, a similar structure appears clearly in the recent
data from the CLAS Collaboration~\cite{Williams:2009yj} (see the
stars in Fig.~\ref{fig-excit}). In our approach the bump structure
is caused by $D_{15}(2080)$. At backward angles, due to the small
$\eta'$ production cross section, it might be difficult to observe
such an enhancement in the excitation functions around $W=2.1$ GeV.

Finally, the total cross section and exclusive cross sections for
several single resonances are illustrated in Fig.~\ref{fig-crp}. The
data can be reasonably well described. The recent data show a small
bump-like structure around $W=2.1$ GeV (see the
stars)~\cite{Williams:2009yj}, which in our approach is due to the
interferences of $D_{15}(2080)$ with other partial waves. Switching
off the contribution of $D_{15}(2080)$, we find that the bump-like
structure disappears (see the dash-dot-dotted curve in the upper
panel of Fig.~\ref{fig-crp}). It should be mentioned that the
bump-like structure around $W=2.1$ GeV was explained by the effects
of $D_{13}(2080)$ and/or $P_{11}2100$ in~\cite{Nakayama:2005ts}.

In Fig.~\ref{fig-crp}, the dominant role of $S_{11}(1535)$ and $u$
channel background can be obviously seen from their exclusive cross
sections, which are much larger than that of other resonances. The
large cross section around the $\eta'$ production threshold mainly
comes from the interferences of $S_{11}(1535)$ and $u$ channel.
Switching off either of them, we find that the cross section will be
underestimated drastically. The calculation shows that both
$S_{11}(1650)$ and $S_{11}(1920)$ have rather small effects on the
cross section around the $\eta'$ production threshold (see the
dotted and dash-dotted curves in the upper panel of
Fig.~\ref{fig-crp}). It should be noted that, although
$S_{11}(1920)$ has a small contribution to the cross section, its
mass favors to be less than $1950$ MeV. Otherwise, we can not
reproduce the present cross sections in the region of $W<2.0$ GeV.
The mass of $S_{11}(1920)$ extracted here is close to that obtained
in Ref.~\cite{Chiang:2002vq}. $S_{11}(1920)$ might correspond to the
$S_{11}(2090)$ listed by the Particle Data Group as a one-star
resonance with a mass varying from 1880 to 2180
MeV~\cite{Nakamura:2010zzi}.

In brief, the $\gamma p\rightarrow \eta' p$ reaction is dominated by
$S_{11}(1535)$ and $u$ channel contributions. The constructive
interference between them accounts for the large cross section near
threshold. $D_{15}(2080)$ plays an important role in the reaction.
It has obvious effects on the angle distributions, and is
responsible for the bump-like structure around $W=2.1$ GeV observed
in the cross section. Weak signal of $S_{11}(1920)$ might be
extracted from the cross section near threshold. The reaction is
less sensitive to the other intermediate states.

\begin{widetext}
\begin{center}
\begin{figure}[ht]
\centering \epsfxsize=15.0 cm \epsfbox{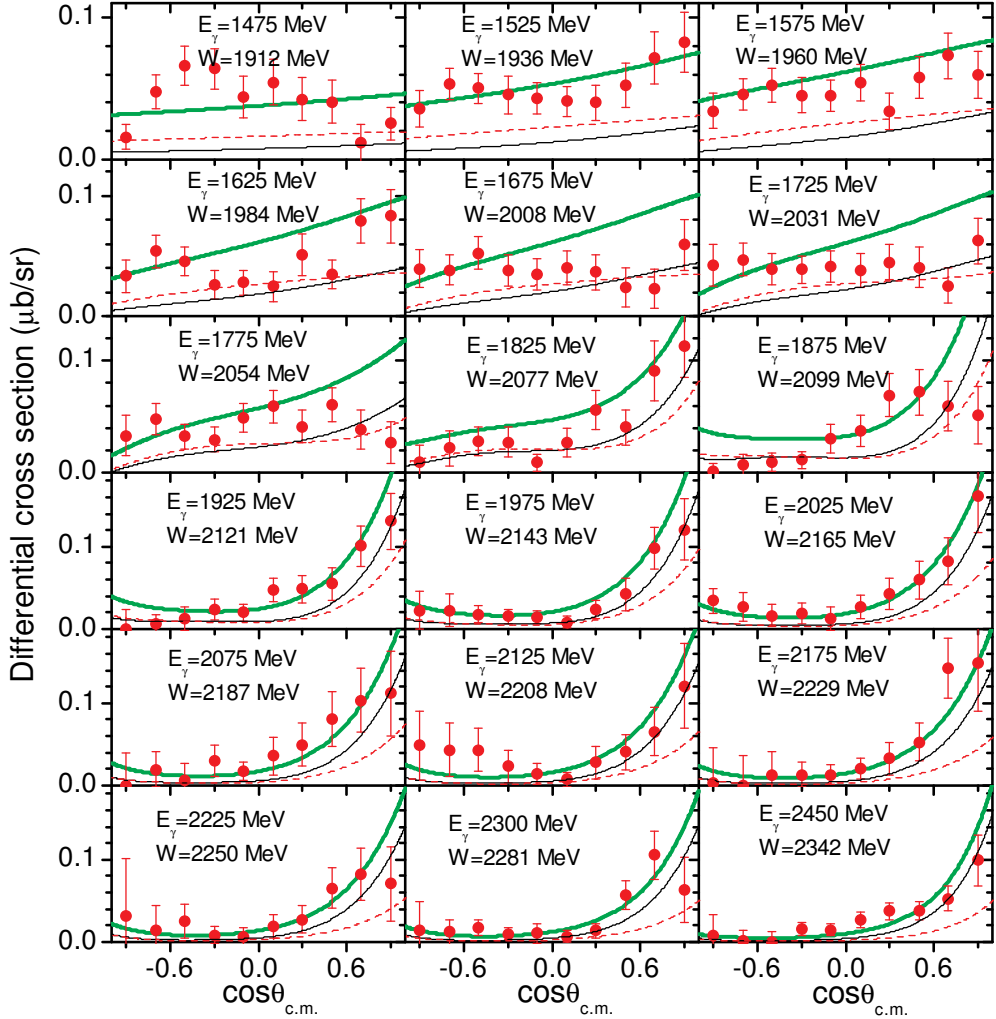} \caption{(Color
online) The differential cross sections for the $\gamma n
\rightarrow \eta'n$ at various beam energies. The data are taken
from~\cite{Jaegle:2010jg} (squares). The bold solid curves stand for
the full model calculations. The thin solid and dotted curves stand
for the results without $S_{11}(1535)$ and background $u$ channel
contributions, respectively. }\label{fig-dfn}
\end{figure}
\end{center}
\end{widetext}

\begin{figure}[ht]
\centering \epsfxsize=8.5 cm \epsfbox{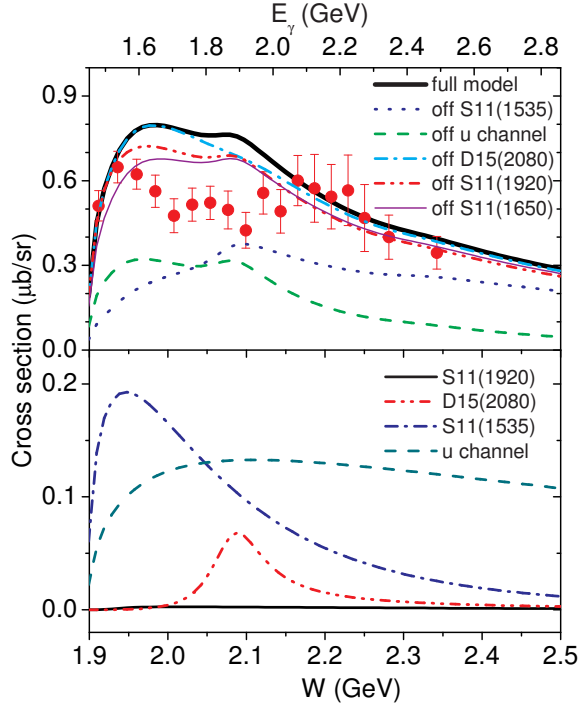} \caption{(Color
online) The cross sections for the $\gamma n\rightarrow \eta' n$
process. The data are taken from~\cite{Jaegle:2010jg}. In the upper
panel the bold solid curve corresponds to the full model result,
while the dotted, thin solid, dash-dot-dotted, dash-dotted, and
dashed curves are for the results by switching off the contributions
from $S_{11}(1535)$, $S_{11}(1650)$, $S_{11}(1920)$, $D_{15}(2080)$
and $u$ channel, respectively. In the lower panel the partial cross
sections for the main contributors are indicated explicitly by
different legends. }\label{fig-crn}
\end{figure}

\begin{figure}[ht]
\centering \epsfxsize=8.5 cm \epsfbox{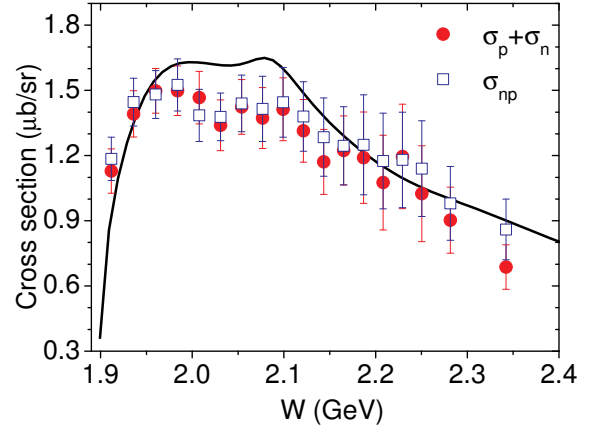} \caption{(Color
online) The data for inclusive quasi-free $\gamma d \rightarrow
np\eta' $ cross section ( $\sigma_{np}$) and the sum of quasi-free
proton and quasi-free neutron cross section
($\sigma_{p}$+$\sigma_{n}$). The solid curve corresponds to our
results of the sum of free proton and free neutron cross section.
}\label{fig-d}
\end{figure}

\subsection{$\gamma n\to \eta' n$}

Recently, the CBELSA/TAPS collaboration had observed the $\gamma
n\to \eta' n$ process for the first time~\cite{Jaegle:2010jg}. The
data had been compared to fits with the NH~\cite{Nakayama:2005ts}
and MAID model~\cite{Chiang:2002vq}. There exists large disagreement
between model fits and the experimental observations. As mentioned
earlier, in $\gamma n\to \eta' n$ states of $[70,^48]$
representation can contribute here while they are forbidden in
$\gamma p\to \eta' p$ by the Moorhouse selection
rule~\cite{Moorhouse:1966jn}. Therefore, we expect that more
information about the $s$-channel resonances can be gained in the
study of $\gamma n\to \eta' n$. For instance, as the only $D_{15}$
state in the first orbital excitations and belonging to $[70, ^48]$,
$D_{15}(1675)$ can only be excited by  $\gamma n$ instead of $\gamma
p$. We also note that in this work the nuclear Fermi motion effects
have been neglected since they are negligible according to the
recent analysis~\cite{Jaegle:2010jg}.

In Fig.~\ref{fig-dfn}, the differential cross sections at various
beam energies have been plotted. It shows that our quark model fits
are in good agreement with the recent CBELSA/TAPS measurements in
the beam energy region $E_{\gamma}> 1.9$ GeV~\cite{Jaegle:2010jg}.
However, in the region $E_{\gamma}< 1.9$ GeV, we can not reproduce
the data well, especially at the forward angles. In this region, our
results are close to the fits of NH model~\cite{Nakayama:2005ts}.

Similar to $\gamma p\to \eta' p$, the differential cross sections
for $\gamma n\to \eta' n$ are governed by the $S_{11}(1535)$ and $u$
channel contributions. Switching off either of them (see thin solid
and dashed curves), we find that the cross sections would be
underestimated significantly. It shows that $S_{11}(1535)$ dominates
near threshold ($E_{\gamma}< 1.9$ GeV), and strongly enhances the
cross section. At higher energies ($E_{\gamma}>2.0$ GeV), the $u$
channel becomes the main contributor in the differential cross
sections. The role of $D_{15}(2080)$ in the $\eta' n$ channel is
similar to that in the $\eta' p$ channel. It slightly enhances the
cross sections at forward angles in the higher energy region
($E_{\gamma}>1.9$ GeV). However, the present data for $\gamma n\to
\eta' n$ seems not precise enough to confirm $D_{15}(2080)$ in the
reaction. Again, we find that the contribution from $P_{13}(1900)$
is negligibly small and might be difficult to identify in the cross
section measurement.

\begin{figure}[ht]
\centering \epsfxsize=8.5 cm \epsfbox{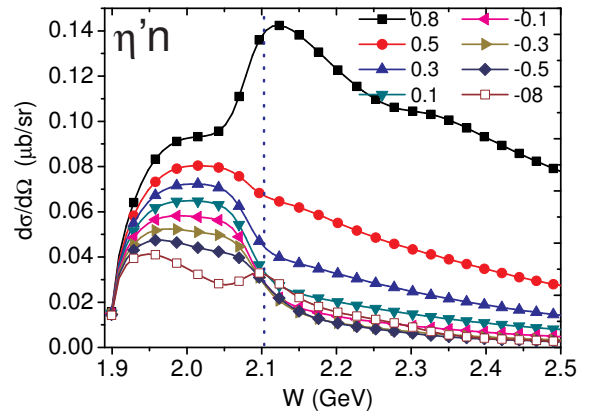} \caption{(Color
online)  The fixed-angle excitation functions for $\gamma
n\rightarrow \eta' n$ as a function of center mass energy $W$ for
eight values of $\cos\theta$, which have been labeled on the
plot.}\label{fig-en}
\end{figure}

In Fig.~\ref{fig-crn}, the total cross section and the exclusive
cross sections of several single resonances are shown. Again, we see
the dominance of $S_{11}(1535)$ and $u$ channel in the cross
sections. Some interfering effects between
$S_{11}(1650)$/$S_{11}(1920)$ and $S_{11}(1535)$ can  be seen near
threshold. There also exist some discrepancies in the low energy
region, i.e. $E_{\gamma}\simeq (1.6\sim 2.0)$ GeV, between our model
results and experimental data. Our model suggests two bump
structures in the total cross section. The first one around $W=1.95$
GeV is mainly caused by $S_{11}(1535)$, while the second around
$W=2.1$ GeV is caused by $D_{15}(2080)$. The
data~\cite{Jaegle:2010jg} seem to show a bump structure around
$W=1.95$ GeV, while the second bump structure around $W=2.1$ GeV can
not be identified due to the large experimental uncertainties.

In Ref.~\cite{Jaegle:2010jg}, the data for the inclusive quasi-free
$\gamma d \rightarrow  np\eta' $ cross section, $\sigma_{np}$, are
also presented. It shows that the $\sigma_{np}$ is nearly equal to
the sum of the free proton ($\sigma_{p}$) and free neutron cross
sections ($\sigma_{n}$). Interestingly, the data indicate two broad
bump structures in the cross section around $W=1.95$ and $W=2.1$
GeV, respectively. To compare with the data we plot our calculations
of $(\sigma_p+\sigma_n)$ in Fig.~\ref{fig-d}, which appears to be
compatible with the data, although the cross section around $W=2.05$
GeV is slightly overestimated. In our approach the second bump
structure in the inclusive quasi-free $\gamma d \rightarrow np\eta'$
cross section is caused by $D_{15}(2080)$. This contribution seems
to be highlighted in $\gamma d \rightarrow  np\eta' $ as a summed-up
effects from both proton and neutron reactions. Further improved
measurement should be able to clarify the under-lying mechanisms
that produces the bump structures.

In Fig.~\ref{fig-en} the excitation functions for $\gamma
n\rightarrow \eta' n$ as a function of the center-of-mass energy $W$
at various angles are plotted. It is sensitive to the presence of
$D_{15}(2080)$ as shown by the drastic enhancement at very forward
angles around $W=2.1$ GeV. This feature is similar to that in
$\gamma p\rightarrow \eta' p$ (see Figs.~\ref{fig-excit} and
~\ref{fig-en}).

\begin{figure}[ht]
\centering \epsfxsize=8.5 cm \epsfbox{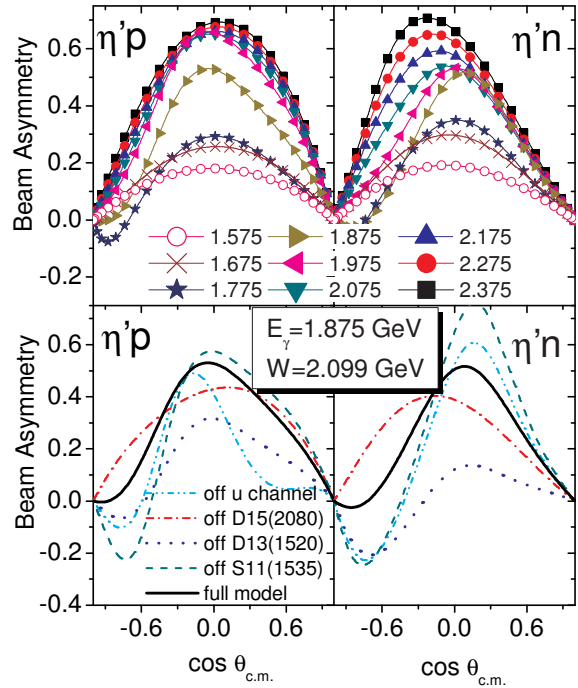} \caption{(Color
online) Predicted beam asymmetries at nine beam energies
($E_{\gamma}=1.575\sim 2.375$ GeV)  for $\gamma p\rightarrow \eta'
p$ and $\gamma n\rightarrow \eta' n$. }\label{fig-ba}
\end{figure}

Polarization observables should be more sensitive to the underlying
mechanisms. In Fig.~\ref{fig-ba}, we plot the polarized beam
asymmetries for $\gamma p\rightarrow \eta' p$ (left) and $\gamma
n\rightarrow \eta' n$ (right), respectively. The beam asymmetries
for both of the precesses are sensitive to $S_{11}(1535)$,
$D_{13}(1520)$, $D_{15}(2080)$ and $u$ channel contributions (see
the bottom of Fig.~\ref{fig-ba}). A sudden change of the beam
asymmetries around $E_\gamma\simeq 1.8$ GeV (i.e. the threshold of
$D_{15}(2080)$) can be seen, which is mainly caused by the
$D_{15}(2080)$. Furthermore, it shows that the beam asymmetry for
$\gamma n\rightarrow \eta' n$ ($\Sigma_n$) is quite similar to that
of $\gamma p\rightarrow \eta' p$ ($\Sigma_p$) up to
$E_{\gamma}\simeq 1.8$ GeV. In this energy region the beam asymmetry
is nearly symmetric in the forward and backward directions. Above
$E_{\gamma}\simeq 1.9$ GeV, obvious differences show up between
$\Sigma_n$ and $\Sigma_p$. It should be noted that the contribution
of $D_{13}(1520)$ does not appear to be significant in the hadronic
model studies. Therefore, experimental measurement of the polarized
beam asymmetries should provide a test for various models.

In brief, $\gamma n\rightarrow \eta' n$ has features similar to
those of $\gamma p\rightarrow \eta' p$. Both reactions are dominated
by $S_{11}(1535)$ and $u$ channel contributions. We predict that
$D_{15}(2080)$ should have significant contributions to $\gamma
n\rightarrow \eta' n$, and the polarized beam asymmetries might be
sensitive to its presence in the transition amplitude. Finally, we
should point out that although $D_{15}(1675)$ has a significant
contribution to $\gamma n\rightarrow \eta n$ process, its
contributions to $\gamma n\rightarrow \eta' n$ is negligible.

\section{Summary}\label{summ}

In this work, we have studied the $\eta'$ photo-production off the
proton and neutron within a chiral quark model. A good description
of the recent experimental data for both processes is achieved. Due
to the similar reaction mechanism for both processes it is
understandable that some similar features exist in both reactions as
manifested in the cross sections, excitation functions and polarized
beam asymmetries.

The large peak of the cross section around threshold for both
processes mainly accounts for the constructive interferences between
$S_{11}(1535)$ and the $u$-channel background. Strong evidence of
$D_{15}(2080)$ has been found in the reactions, with which we can
naturedly explain the following recent high-statistics observations
for the $\gamma p\rightarrow \eta' p$ reaction from the CLAS
Collaboration: (i) the sudden change of the shape of the
differential cross section around $E_{\gamma}=1.8$ GeV, (ii) the
bump-like structure in the total cross section around $W=2.1$ GeV
($E_{\gamma}\simeq 1.9$ GeV), and (iii) the peak around $W=2.1$ GeV
in the excitation functions at very forward angles. Furthermore,
$D_{15}(2080)$ also accounts for the bump-like structure at
$W\simeq2.1$ GeV in the inclusive quasi-free $\gamma d \rightarrow
np\eta' $ cross section measured by CBELSA/TAPS.

$S_{11}(1920)$ seems to be needed in the reaction, with which the
total cross section near threshold for $\gamma p\rightarrow \eta' p$
is improved slightly. However, the differential cross sections,
excitation functions, and beam asymmetries are not sensitive to
$S_{11}(1920)$. To confirm $S_{11}(1920)$, more accurate
observations are needed.

Furthermore, it should be mentioned that the polarized beam
asymmetries are found to be sensitive to $D_{13}(1520)$, although
its effects on the differential cross sections and total cross
sections are negligible. There is no obvious evidence of the $P$-,
$D_{13}$-, $F$-, and $G$-wave resonances with a mass around 2.0 GeV
in the reactions.

To better understand the physics in the $\gamma p\rightarrow \eta'
p$ and $\gamma n\rightarrow \eta' n$ reactions, we expect more
accurate measurements of the following observables for both of the
processes: (i) the total cross section in the energy region
$E_{\gamma}\simeq (1.55\sim 2.1)$ GeV, (ii) the fixed-angle
excitation functions at very forward angles from threshold up to
$W\simeq 2.3$ GeV, (iii) the differential cross sections in the
energy region $E_{\gamma}\simeq (1.6\sim 1.9)$ GeV, and (iv) the
beam asymmetries in the energy region $E_{\gamma}\simeq (1.6\sim
2.0)$ GeV.

\section*{  Acknowledgements }

The authors thank B. Krusche for providing us the data of $\eta'$
photoproduction off quasi-free nucleons. This work is supported, in
part, by the National Natural Science Foundation of China (Grants
10775145, 11075051 and 11035006), Chinese Academy of Sciences
(KJCX2-EW-N01), Ministry of Science and Technology of China
(2009CB825200), the Program for Changjiang Scholars and Innovative
Research Team in University (PCSIRT, No. IRT0964), the Program
Excellent Talent Hunan Normal University, and the Hunan Provincial
Natural Science Foundation (11JJ7001).


\end{document}